\documentclass[twocolumn,showpacs,preprintnumbers,amsmath,amssymb,prb]{revtex4-1}

\usepackage{graphicx}% Include figure files
\usepackage{dcolumn}% Align table columns on decimal point
\usepackage{bm}% bold math
\usepackage{mathrsfs}
\usepackage{color}
\usepackage{amsmath}
\usepackage{amssymb}

\begin{document}

\title{Parity qubits and poor man's Majorana bound states in double quantum dots}

\date{\today}
\author{Martin Leijnse}
\author{Karsten Flensberg}
\affiliation{Center for Quantum Devices, Niels Bohr Institute, University of Copenhagen, Universitetsparken 5, DK-2100 Copenhagen, Denmark}

\begin{abstract}
We study a double quantum dot connected via a common superconducting lead and show that this system can be tuned to host one Majorana bound state (MBS) on each dot. We call them "poor man's Majorana bound states" since they are not topologically protected, but otherwise share the properties of MBS formed in topological superconductors. We describe the conditions for the existence of the two spatially separated MBS, which include breaking of spin degeneracy in the two dots, with the spins polarized in different directions. Therefore, we propose to use a magnetic field configuration where the field directions on the two dot form an angle. By control of this angle the cross Andreev reflection and the tunnel amplitudes can be tuned to be approximately equal, which is a requirement for the formation of the MBS. We show that the fermionic state encoded in the two Majoranas constitutes a parity qubit, which is non-local and can only be measured by probing both dots simultaneously. Using a many-particle basis for the MBS, we discuss the role of interactions and show that interactions between electrons on different dots always shift the condition for degeneracy. We also show how the MBS can be probed by transport measurements and discuss how the combination of several such double dot systems allows for entanglement of parity qubits and measurement of their dephasing times. 
\end{abstract}
\maketitle
\section{Introduction}

It is well known that as a result of Andreev reflection at the normal metal--superconductor interface, the pairing interaction present in superconductors can be transferred to normal metals and semiconductors by the so-called proximity effect. Recently, this effect has been suggested as a way to induce a $p$-wave pairing \cite{Ivanov2001, Fu2008, Sato2009,Potter2010} needed to create interesting topological states associated with Majorana bound states (MBS) in conventional semiconductors by a combination of spin-orbit coupling and Zeeman splitting induced by external magnetic fields, \cite{Sau2010,Alicea2010,Lutchyn2010,Oreg2010} or by spatially varying magnetic fields without spin-orbit interaction. \cite{Kupferschmidt2011,Kjaergaard2012,EggerFlensberg2012} See Refs.~\onlinecite{BeenakkerMajoranaReview,AliceaMajoranaReview,LeijnseMajoranaReview} for recent reviews on this rapidly developing field. MBS are interesting because of the potential use as elements in a topological quantum computing architecture.\cite{Nayak2008} Even though topological manipulations of MBS do not allow an universal set of gates, it could have advantages for a restricted set of operations or for storage of quantum information.

Another usage of induced pairing is the so-called Cooper pair splitter, where Cooper pairs are split through cross Andreev reflection,
which gives a source of entangled electrons because of the singlet nature of the Cooper pairs.\cite{Lesovik2001} This idea was further theoretically developed to include a quantum dot in each arm of the beam splitter,\cite{Recher2001,Bouchiat2003} a geometry which was later realized using carbon nanotubes\cite{Herrmann2010} and nanowires.\cite{Hofstetter2009}

The ideas of cross Andreev reflection and induced $p$-wave superconductivity in a semiconductor system were combined in a recent proposal by Sau and Das Sarma, see Ref.~\onlinecite{SauDasSarma}. A series of quantum dots, spin split by a magnetic field, but with non-collinear spin arrangements due to spin-orbit coupling, make a direct mapping of the Kitaev model\cite{Kitaev2001} onto an engineered quantum dot system. Since quantum dot technology is well established, this proposal has clear advantages over others relaying on particular material properties. 

Here we consider a very simple system, sketched in Fig.~\ref{fig:1}, consisting of two quantum dots tunnel coupled to a common $s$-wave superconductor. In addition, large non-collinear magnetic fields are applied to the dots. This setup allows for splitting of a Cooper pair into the dots (the split electrons are, however, not entangled because of the dot spin polarizations).
This, in turn, creates a possibility to induce a $p$-wave pairing potential between electrons residing in the dots.
The angle of the dot magnetic fields gives a handle on the ratio of the normal tunneling and cross Andreev tunneling, allowing for a simple tuning into the interesting regime with MBS localized on the dots.

The MBS in our setup are not protected to the same degree as topological states in $p$-wave superconductors. The system is nonetheless useful for testing theoretical predictions of, e.g., resonant Andreev reflection\cite{Law2009,Flensberg2010}, non-local teleportation-like phenomena \cite{Fu2010}, and measurements of the lifetime of the non-local state carrying information about the parity of the two Majorana states. Given the small Hilbert space of the system we can explicitly study the influence of interactions by expressing the Majorana states in a many-body language. We also calculate the transport properties with the double dot system tunnel-coupled to a normal metallic probe, and furthermore discuss experimental setups that would allow a determination of the dephasing and lifetimes of qubits based on the parity of the Majorana states. 

The paper is organized as follows: In Sec.~II we set up the model and calculate the conditions for having a set of MBS. We also investigate the sensitivities to fluctuations of the various parameters, as well as the effects of inter-dot electron-electron interaction. In Sec.~III, we show the expected tunneling characteristic, and finally Sec.~IV is concerned with entanglement and decoherence of the parity qubits.

\section{Double dot model and Majorana states}
\begin{figure}[t!]
  	\includegraphics[height=0.44\linewidth]{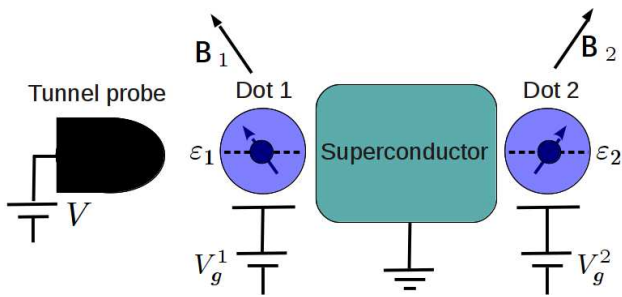}	
	\caption{\label{fig:1}(Color online) Sketch of setup. Two quantum dots are coupled to each other via a superconductor, which mediates both 
	normal tunneling between the dots and cross Andreev reflection. Each dot has only one level close to the chemical potential 
	of the superconductor (energies $\varepsilon_1$ and $\varepsilon_2$), which are controlled by the 
	gate voltages $V_g^1$ and $V_g^2$. The dots are fully spin-polarized, but in different directions because of the inhomogeneous 
	magnetic field ($\mathbf{B}_1$ at dot 1 and $\mathbf{B}_2$ at dot 2).
	The normal metal tunnel probe, coupled to dot 1, is used in Sec.~\ref{sec:tunnel} to detect the MBS.}
\end{figure}

We consider a double quantum dot connected via a common superconducting lead. The width of the superconductor is smaller than the superconducting coherence length, which allows cross Andreev reflection involving electrons on different dots. Furthermore, electrons can tunnel via the superconductor from one dot to the other, involving virtual occupation of quasiparticle states above the gap. The geometry is illustrated in Fig.~\ref{fig:1}.
The magnetic fields on  the two dots, $\mathbf{B}_1$ and $\mathbf{B}_2$, make an angle $\varphi$. We will assume the Zeeman splitting to be larger than temperature, which allows us to consider only one state in each dot. The amplitude for tunneling between the dots therefore depends on the angle $\varphi$ as $t=t_0\cos(\varphi/2)$, where $t_0$ is the tunneling amplitude for parallel fields. Similarly, the cross Andreev reflection induces a pair coupling between electrons in the two dots which also depends on the angle. However, since we consider a standard $s$-wave superconductor, the pairing is maximal for antiparallel spin polarizations, and is given by $\Delta=\Delta_0 \sin(\varphi/2)$. 
Thus, $\varphi$ provides a way to tune the ratio $t/\Delta$, which will be crucial to engineer the appropriate conditions for MBS.

The model Hamiltonian is 
\begin{equation}\label{H}
  H = \varepsilon_1^{{}} n_1 + \varepsilon_2^{{}} n_2 + t d_1^{\dagger} d_2^{{}} + \Delta d_1^{\dagger} d_2^{\dagger} + h.c.,
\end{equation}
where $n_{i} = d_{i}^{\dagger} d_{i}^{{}}$ is the occupation operator for dot $i = 1,2$, and where the onsite energies, $\varepsilon_{i}$, are measured relative to the chemical potential of the superconductor.

\subsection{Single-particle formulation}

We start by studying the Hamiltonian \eqref{H} within the Nambu formalism. Using the basis $\Psi=(d_1^{{}},d_2^{{}},d_1^\dagger,d_2^\dagger)$, Eq.~\eqref{H} can be written
\begin{equation}\label{HNambu}
  H = \frac12 \Psi^\dagger {h}\Psi+\frac12(\varepsilon_1+\varepsilon_2),
\end{equation}
with
\begin{equation}\label{HNambu1st}
  {h}= \left(\begin{array}{cccc}
     \varepsilon_1 & t & 0 & \Delta\\
     t & \varepsilon_2 & - \Delta & 0\\
     0 & - \Delta & - \varepsilon_1 & - t\\
     \Delta & 0 & - t & - \varepsilon_2
   \end{array}\right),
\end{equation}
where we have chosen both $t$ and $\Delta$ real. The eigenvalues, $\lambda$, of the Hamiltonian matrix, ${h}$, fulfill
\begin{equation}\label{Es}
  \lambda^2= \varepsilon_+^2+\varepsilon_-^2+t^2+\Delta^2+2\eta\sqrt{(\varepsilon_+^2+\Delta^2)(\varepsilon_-^2+t^2)},
\end{equation}
where $\varepsilon_\pm=(\varepsilon_1\pm\varepsilon_2)/2$, and $\eta=\pm1$. 

Since we are searching for MBS, we seek zero energy solutions to Eq.~(\ref{Es}).
Therefore, we set $\eta=-1$ and see that for $t=\pm\Delta$, the eigenenergy vanish for either $\varepsilon_1$ or $\varepsilon_2$ 
being equal to zero. 
We start by investigating the ''sweet spot'' (we show below that the MBS are quadratically protected in this point), where in addition to $t=\Delta$, both dot levels are aligned with the chemical potential of the superconductor, $\varepsilon_{1}= \varepsilon_2 = 0$. Here, the solutions to the Bogoliubov-de Gennes equations, $h \psi_i = E_i \psi_i$, are
\begin{subequations}\label{eigenstates}
\begin{align}
\label{eigenstates1}
  \psi_1 &= \frac{1}{\sqrt{2}} \left(\begin{array}{cccc}
    1 & 0 & 1 & 0
  \end{array}\right)^T,\quad E_1= 0,  \\
\label{eigenstates2}
  \psi_2 &= \frac{i}{\sqrt{2}} \left(\begin{array}{cccc}
    0 & 1 & 0 & - 1
  \end{array}\right)^T,\quad  E_2 = 0,   \\
\label{eigenstates3}
  \psi_3 &= \frac{1}{2} \left(\begin{array}{cccc}
    - 1 & 1 & 1 & 1
  \end{array}\right)^T,\quad  E_3 = - 2 t,   \\
\label{eigenstates4}
  \psi_4 &= \frac{1}{2} \left(\begin{array}{cccc}
    1 & 1 & - 1 & 1
  \end{array}\right)^T,\quad  E_4 = 2 t.
\end{align}
\end{subequations}
The corresponding second quantization operators are given by $\gamma_i = \Psi \cdot \psi_i$. 
For the two zero energy states we then find $\gamma_1 = (d_1^{{}} + d_1^{\dagger})/\sqrt{2}$ and $\gamma_2 = i(d_2^{{}} - d_2^{\dagger})/\sqrt{2}$.
These are clearly Hermitian, $\gamma_{1,2}^{{}}=\gamma_{1,2}^\dagger$, and therefore describe MBS. 
Furthermore, the two MBS are spatially isolated since each zero energy state is completely localized on one of the dots.

If we let only one dot level move away from 0, say $\varepsilon_1 \neq 0$, the two low energy states remain doubly degenerate. The corresponding eigenstates of $h$ are in this case
\begin{subequations}\label{eigenstates_change_e1}
\begin{align}
\label{eigenstates1_change_e1}
  \psi_1 &=\frac{1}{A \sqrt{2}} \left(\begin{array}{cccc}
    1 & - \delta & 1 & - \delta
  \end{array}\right)^T, \quad E_1 = 0,  \\
\label{eigenstates2_change_e1}
  \psi_2 &= \frac{1}{\sqrt{2}} \left(\begin{array}{cccc}
    0 & 1 & 0 & - 1
  \end{array}\right)^T, \quad E_2 = 0 ,
\end{align}
\end{subequations}
with $\delta=\varepsilon_1/2t$ and $A=\sqrt{1+\delta^2}$. Thus, both $\psi_1$ and $\psi_2$ are still MBS, but while $\psi_2$ is completely localized on dot 2, $\psi_1$ also has a component ($\propto
\frac{\varepsilon_1}{t}$) on dot 2.

To study the sensitivity of the zero energy states, we expand up to second order in the onsite energies, $\varepsilon_{1,2}$, at the point where $t=\pm\Delta$
\begin{equation}\label{Eexpand}
E_{1,2}=\pm \frac{\varepsilon_{1}\varepsilon_{2}}{2\Delta}\left[1+\mathcal{O}\left( \left( \frac{\varepsilon_{1,2}}{\Delta} \right)^2 \right) \right],
\end{equation}
which shows that the zero energy solutions are ``protected" against small deviations to linear order in the onsite energies. In contrast, deviations away from $t=\Delta$ result in
\begin{equation}\label{texpand}
E_{1,2} = \pm (|\Delta|-|t|).
\end{equation}
Thus, there is no protection against such devations. However, one can still find a condition for zero modes even for $|t|\neq|\Delta|$ by adjusting the onsite energies, but these will not be quadratically protected.

In general, the diagonalized Nambu Hamiltonian, $h$, in Eq.~\eqref{HNambu} has eigenvectors $\psi_{i}$ with eigenenergies $E_i$, where $i=1,\ldots4$.
Due to electron-hole symmetry the eigenvalues come in pairs with energies $\pm E$. Choosing $E_1=-E_2$, $E_3=-E_4$, we can write the Hamiltonian \eqref{H} as
\begin{equation}\label{Hdia}
  H=|E_1|\beta^\dagger_1\beta_1^{{}}+|E_3|\beta^\dagger_3\beta_3^{{}}+\frac12(\varepsilon_1+\varepsilon_2-|E_1|-|E_3|).
\end{equation}
The ground state energy is thus given by $E_g=\frac12(\varepsilon_1+\varepsilon_2-|E_1|-|E_3|)$, which for the special case $t=\Delta$ and $\varepsilon_{1,2}=0$ becomes $E_g=-|t|$. The ground state is thus two-fold degenerate, because $E_1=0$. The degeneracy corresponds to the occupation of the fermion formed by $f=(\gamma_1-i\gamma_2)/2$, with the two MBS residing on the two dots. Below we study this conclusion in a many-particle formulation.

\subsection{Many-particle formulation of Majorana fermions}
In the basis $\{|00 \rangle, |10 \rangle, |01 \rangle, |11\rangle\}$ of number states $|n_1 n_2\rangle$ (where $|11 \rangle \equiv d_1^{\dagger} d_2^{\dagger} |00 \rangle$ fixes the choice of sign), the many-particle version of the Hamiltonian \eqref{H} becomes
\begin{equation}\label{Hmany}
   H = \left(\begin{array}{cccc}
     0 & 0 & 0 & \Delta\\
     0 & \varepsilon_1 & t & 0\\
     0 & t & \varepsilon_2 & 0\\
     \Delta & 0 & 0 & \varepsilon_1 + \varepsilon_2
   \end{array}\right),
\end{equation}
where we again have chosen both $t$ and $\Delta$ real. The states $|01 \rangle$ and
$|10 \rangle$ couple via the normal tunneling, $t$, while $|00 \rangle$ and
$|11 \rangle$ couple via the cross Andreev reflection, $\Delta$.

We saw above that MBS exist in the special limit $\varepsilon_1 = \varepsilon_2 = 0, t = \pm
\Delta$. Taking $t = \Delta$, the eigenstates are:
\begin{subequations}\label{sweetspotmany}
\begin{align}
\label{sweetspotmany1}
  | \alpha_e \rangle &= \frac{1}{\sqrt{2}} (|00 \rangle + |11 \rangle),\quad
  E_{\alpha_e} = t,  \\
\label{sweetspotmany2}
  | \alpha_o \rangle &= \frac{1}{\sqrt{2}} (|10 \rangle + |01 \rangle),\quad
  E_{\alpha_o} = t,  \\
\label{sweetspotmany3}
  | \beta_e \rangle &= \frac{1}{\sqrt{2}} (|00 \rangle - |11 \rangle),\quad
  E_{\beta_e} = - t, \\
\label{sweetspotmany4}
  | \beta_o \rangle &= \frac{1}{\sqrt{2}} (|10 \rangle - |01 \rangle),\quad
  E_{\beta_o} = - t.
\end{align}
\end{subequations}

There are two degenerate pairs of eigenstates ($\alpha$ and $\beta$), with one of the states in each pair having even ($e$) and the other odd ($o$) fermion number parity. In the MBS language, these two states are eigenstates of the number operator corresponding to the Dirac fermion operator made out of the two MBS. To see this,  we use the non-local fermion $f = (\gamma_1 - i \gamma_2)/2$ with the corresponding occupation $n = f^{\dagger} f = (1 - i \gamma_1\gamma_2)/2$, which in terms of the original $d$ fermions becomes
\begin{equation}\label{nd}
 n = \frac{1}{2} (1 + d_1^{\dagger} d_2 - d_1 d_2^{\dagger} + d_1 d_2 -
   d_1^{\dagger} d_2^{\dagger}).
\end{equation}
Acting with this number operator on the eigenstates of the many-particle Hamiltonian, we then have
\begin{equation}
  n| \alpha_e \rangle  = n| \beta_e \rangle  = 0, \quad  n| \alpha_o \rangle  = | \alpha_o \rangle,\quad n| \beta_o \rangle = | \beta_o \rangle.
\end{equation}
Thus, $| \alpha_e \rangle$, $| \beta_e \rangle$ and $| \alpha_o \rangle$, $| \beta_o \rangle$ are eigenstates of the number operator corresponding to the non-local fermion $f$, with eigenvalues 0 and 1, respectively. The two-fold degeneracy of the ground state thus corresponds to an even or odd number of fermions in the total system consisting of the superconductor and quantum dots. Moreover, when operating on a ground state, the Majorana operators flip the parity, for example
\begin{equation}\label{gamma1ae}
  \gamma_1| \alpha_e \rangle=\frac{1}{\sqrt{2}} (d_1^{{}}+d_1^\dagger)| \alpha_e\rangle = | \alpha_o \rangle.
\end{equation}

\subsection{Non-locality of parity measurements}

The non-local nature of the two degenerate sweet-spot ground states 
[Eqs.~\eqref{sweetspotmany1} and~\eqref{sweetspotmany2} for $t<0$ and Eqs.~\eqref{sweetspotmany3} and~\eqref{sweetspotmany4} for $t>0$]  
has the consequence that one cannot
distinguish between them by local measurements on a single dot. A measurement of the charge on one dot,
$Q_{1, 2} = - e \langle n_{1, 2} \rangle_{}$, gives the same result for the even and odd parity states, $ \langle \beta_e |n_{1, 2} | \beta_e \rangle 
=  \langle \beta_o |n_{1, 2} | \beta_o \rangle  = \frac{1}{2}$ (and the same for the $\alpha$-states). 
In fact, that the parity states cannot be distinguished by a local measurement is clear since they are maximally entangled Bell states 
in terms of the dot charges, see Eq.~(\ref{sweetspotmany}). 
However, a measurement of the fluctuations of the \emph{total} charge does reveal the state,
since $  \langle \beta_e | (n_1 + n_2)^2 | \beta_e \rangle  = 2$, while $ \langle \beta_o | (n_1 + n_2)^2 | \beta_o \rangle  = 1$. Detection of the fluctuations could be done by the having a non-linear charge detector, e.g., a single-electron transistor, capacitively coupled to both dots.

The non-locality is, however, destroyed if the system is tuned away from the sweet spot. To show this we consider the situation where $|t|\neq |\Delta|$, while tuning to a degeneracy point by setting $\varepsilon_1=(t^2-\Delta^2)/\varepsilon_2$. In this case the ratio of the occupations for the even/odd states of dot 2 become (to lowest order in $|t|-|\Delta|$):
\begin{equation}\label{n1n2}
    \frac{ \langle n_{2} \rangle_e}{ \langle n_{2} \rangle_o}\approx 1-\frac{\sqrt{4\Delta^2+\varepsilon_2^2}+\varepsilon_2}{\Delta\varepsilon_2} (|t|-|\Delta|),
\end{equation}
which shows that the non-locality of the determination of the dot occupations is gradually destroyed as one moves away from the sweet spot.

\subsection{The influence of interdot interaction}

An interaction between the charge on the two dots corresponds to a term in the Hamiltonian given by
\begin{equation}\label{HU}
    H_U=Un_1n_2,
\end{equation}
which in the many-particle basis of Eq.~\eqref{Hmany} becomes
\begin{equation}\label{HmanyU}
   H +H_U= \left(\begin{array}{cccc}
     0 & 0 & 0 & \Delta\\
     0 & \varepsilon_1 & t & 0\\
     0 & t & \varepsilon_2 & 0\\
     \Delta & 0 & 0 & \varepsilon_1 + \varepsilon_2+U
   \end{array}\right).
\end{equation}
The two lowest eigenenergies are then
\begin{subequations}
\begin{align}
  E_1 &= \varepsilon_+ - \sqrt{(\varepsilon_+)^2 + t^2},\\
  E_2 &=  \frac12 \left( 2 \varepsilon_+ + U - \sqrt{(2 \varepsilon_+ + U)^2+4\Delta^2}\right).
\end{align}
\end{subequations}
The situation with a finite $U$ is similar to the case discussed above, when $t$ and $\Delta$ are tuned away from the sweet spot. 
One can tune the system into a situation with two MBS, but their existence is not quadratically protected.
The two energies are degenerate only if
\begin{equation}\label{Ucond}
    U=\frac{\left(\Delta^2-t^2+\varepsilon_1\varepsilon_2\right)\left(\varepsilon+2\sqrt{t^2 + \varepsilon_-^2}\right)}{2(t^2-\varepsilon_1\varepsilon_2)}.
\end{equation}
Moreover, the fermion associated with the parity of the ground state is no longer fully non-local in nature, because one can determine the state by a measurement of the charge on one dot only, similar to the situation for the non-interacting case away from the sweet spot. 

In an experiment, we do not expect it to be particularly problematic to achieve $U \approx 0$ since the superconductor efficiently screens the charge on one dot as seen from the other. Note that local (intra-dot) electron--electron interactions are irrelevant since both dots are fully spin polarized (only single occupancy is allowed).

\section{Detecting the Majorana states by tunnel spectroscopy\label{sec:tunnel}}

A simple way to detect the MBS is by tunnel spectroscopy. We consider a normal metallic electrode tunnel coupled to dot 1, 
see Fig.~\ref{fig:1}. 
We want to determine the current flowing into the grounded superconductor from the normal electrode, as a function of the applied bias voltage.
We assume that the normal electrode is weakly coupled to dot 1, such that the entire voltage drop takes place at the normal electrode--quantum dot 
tunnel barrier. The current is determined by the Andreev reflection amplitude $a(\omega)$ as 
\begin{equation}\label{I}
    I=\frac{2e}{h}\int_{-\infty}^\infty d\omega |a(\omega)|^2[f(\omega-eV)-f(\omega+eV)],
\end{equation}
where $f(\omega)$ is the Fermi-Dirac distribution of the normal electrode. 
\begin{figure*}[t!]
  	\includegraphics[height=0.43\linewidth]{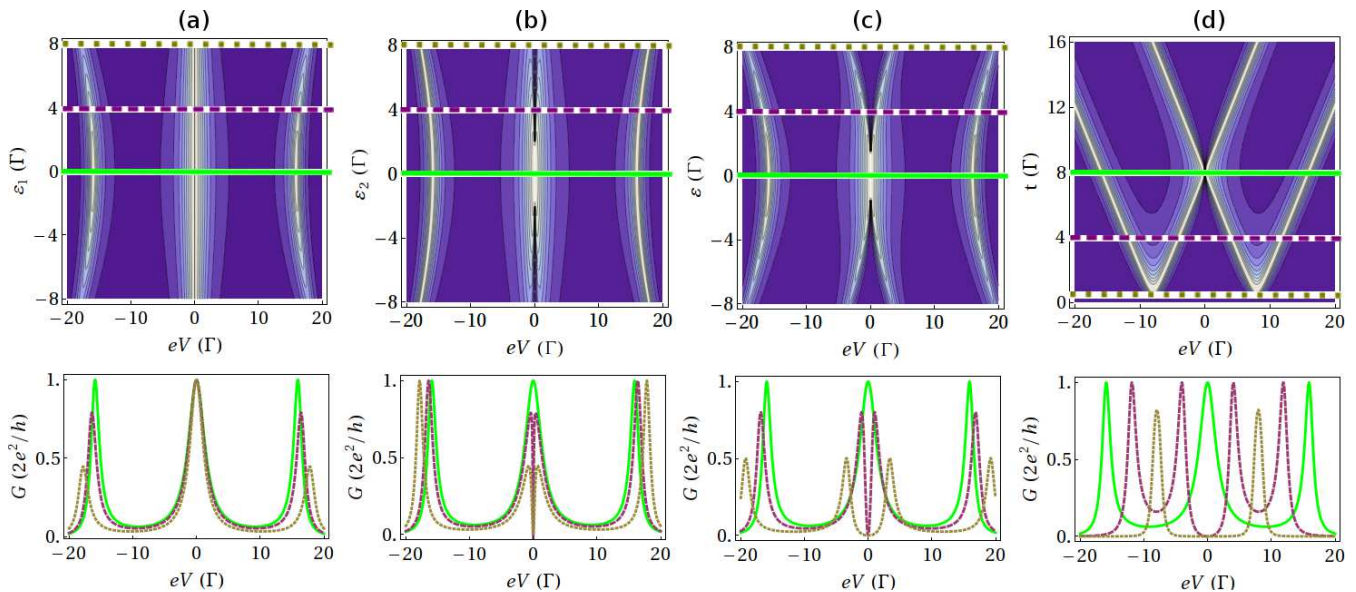}	
	\caption{\label{fig:2}(Color online) Calculated differential conductance, $G = dI/dV$, at zero temperature. 
	Upper panel: $G$ plotted on a color scale as a function of bias voltage and one more parameter, while all other parameters are fixed at their 
	sweet spot values, with $t = \Delta = 8\Gamma$. Lower panel: $G$ as a function of bias voltage along three 
	different horizontal cuts in the corresponding upper panel conductance map (the position being marked there with the same color and linestyle as 
	the corresponding curve in the lower panel). 
	(a) $\varepsilon_1$ is varied, cuts in lower panel are at $\varepsilon_1 = 0$ (green solid curve), $\varepsilon_1 = 4 \Gamma$ (magenta dashed curve), 
	and $\varepsilon_1 = 8 \Gamma$ (brown fine-dashed curve).
	(b) $\varepsilon_2$ is varied, cuts in lower panel are at $\varepsilon_2 = 0$ (green solid curve), $\varepsilon_2 = 4 \Gamma$ (magenta dashed curve), 
	and $\varepsilon_2 = 8 \Gamma$ (brown fine-dashed curve).
	(c) $\varepsilon = \varepsilon_1 = \varepsilon_2$ is varied, cuts in lower panel are at $\varepsilon = 0$ (green solid curve), $\varepsilon = 4 \Gamma$ 
	(magenta dashed curve), 
	and $\varepsilon = 8 \Gamma$ (brown fine-dashed curve).
	(d) $t$ is varied, cuts in lower panel are at $t = \Delta = 8 \Gamma$ (green solid curve), $t = 4 \Gamma$ (magenta dashed curve), 
	and $t = \Gamma / 2$ (brown fine-dashed curve).}
\end{figure*}

The amplitudes for normal reflection, $r(\omega)$, and Andreev reflection, $a(\omega)$, follow from the scattering matrix 
\begin{equation}\label{Sdef}
    S=\left(
        \begin{array}{cc}
          r & a \\
          a^* & r^* \\
        \end{array}
      \right).
\end{equation}
In the wide-band limit, the $S$-matrix is given by 
\begin{equation}\label{S}
    S=1+2i\pi W[h-\omega+i\pi W^\dagger W]^{-1}W^\dagger
\end{equation}
where $W$ is the coupling matrix describing the coupling between the normal electrode and the dot system. With a tunnel 
coupling only to dot 1, it is given by
\begin{equation}\label{W}
    W=\sqrt{\frac{\Gamma}{2\pi}}
    \left(
      \begin{array}{cccc}
        1 & 0 & 0 & 0 \\
        0 & 0 & -1 & 0 \\
      \end{array}
    \right),
\end{equation}
where $\Gamma$ is the width of the dot level due to the tunnel coupling.
 
In Fig.~\ref{fig:2}, we show the calculated zero-temperature differential conductance, $G = dI/dV$. In each subfigure, (a)--(d), 
the upper panel shows $G$ on a color scale plotted as a function of $V$ and one more parameter, while all other parameters are kept 
fixed at their sweet-spot values. The lower panel shows $G$ as a function of $V$ along three different horizontal cuts in the corresponding 
upper panel conductance map. 
At the sweet spot, found along a horizontal line through the center of each upper panel conductance map and represented by the green conductance curve in 
each lower panel, we find a peak of height $2e^2/h$ centered at $V=0$ and broadened by $\Gamma$. 
This is a well-known result for tunneling into a localized Majorana bound 
state.~\cite{Law2009,Flensberg2010} The additional states in Eq.~(\ref{eigenstates}) give rise to conductance peaks at $\pm 2t$. 

In Fig.~\ref{fig:2}(a), we let $\varepsilon_1$ vary away from the sweet-spot value ($\varepsilon_1 = 0$). As discussed above, this does not remove the 
zero-energy states, but only moves some of the weight of the Majorana wavefunction $\psi_1$ from dot 1 to dot 2, see Eq.~(\ref{eigenstates_change_e1}). 
As a result, the zero-bias peak remains, but with a somewhat reduced width, related to the reduced weight of the Majorana wavefunction on 
dot 1. The finite bias conductance signatures show a stronger dependence on $\varepsilon_1$, moving to higher $V$ and being reduced in height.
Varying instead $\varepsilon_2$ away from the sweet spot, as in Fig.~\ref{fig:2}(b), leads to a qualitatively different conductance map.
Now some of the Majorana wavefunction localized on dot 2, $\psi_2$, "spills over" into dot 1, enabling tunneling from the normal electrode into 
both MBS. Tunneling into the two modes interfere destructively, leading to a sharp dip in the conductance centered at $V = 0$ 
(if the temperature is too large to resolve this dip, the decreased height of the zero bias peak still provides a transport signature of 
$\varepsilon_2 \neq 0$). 
Note, however, that the zero-energy states remain intact.
In Fig.~\ref{fig:2}(c), we simultaneously move both dot levels away from zero, setting $\varepsilon_1 = \varepsilon_2 = \varepsilon$. This introduces 
a splitting of the zero-energy states in quadratic order, see Eq.~(\ref{Eexpand}), and therefore also a splitting of the zero bias conductance peak. 
Varying $t$ away from $\Delta$, as in Fig.~\ref{fig:2}(d), introduces a linear splitting of the zero energy states, Eq.~(\ref{texpand}), 
and of the corresponding 
conductance peak. For very small $t$ (or very small $\Delta$), the conductance is suppressed. Below the superconducting gap, the only way to move 
electrons into or out of the 
superconductor, and thereby to or from ground, is through cross Andreev reflection. However, this can only happen when both dots are 
either empty or full, necessitating normal tunneling since the normal electrode is only coupled to dot 1.

The results in Fig.~\ref{fig:2} show that tunnel spectroscopy can indeed be used to detect the MBS. Moreover, any parameter 
being tuned away from the sweet spot results in a clear conductance signature. Therefore, continuously monitoring the conductance 
spectrum provides a guide for an experimentalist navigating through the parameter space ($\varepsilon_1$, $\varepsilon_2$, and $t/\Delta$) 
towards the sweet spot.
 
\section{Parity qubits: Entanglement and coherence times}
The fermionic two-level system spanned by the MBS can be thought of as a parity qubit, the state of which can be read out 
via the parity measurements discussed above. 
However, there appears to be no feasible way to create superposition states of an isolated parity qubit, i.e., to rotate it away from the north or south pole on the block sphere (even or odd parity).
Controlled addition or removal of an electron changes the parity between even and odd (flips the qubit between north and south pole), 
but rotations by other angles would require adding or removing a fractional charge. 

\begin{figure}[h!]
  	\includegraphics[height=0.75\linewidth]{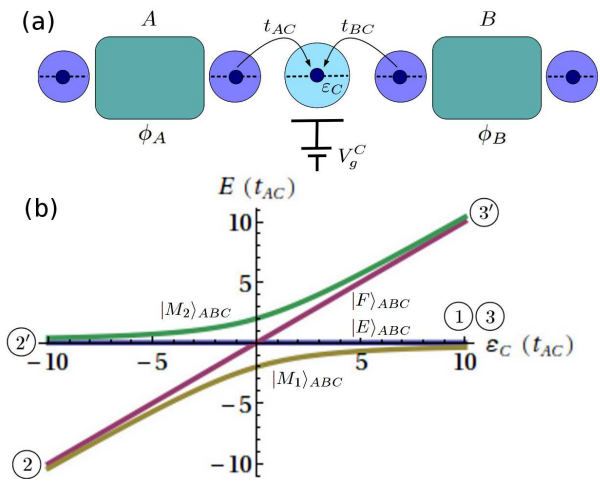}	
	\caption{\label{fig:3}(Color online) (a) Sketch of setup with two double dots, tunnel coupled with amplitudes $t_{AC}$ and $t_{BC}$ to 
	an additional dot ($C$). Dot $C$ has a single orbital with energy $\varepsilon_C$, which is controlled by the gate voltage $V_g^C$.
	(b) The eigenenergies of the even total parity sector of Eq.~(\ref{H_ABC}) for $t_{AC} = t_{BC}$, plotted as a function of $\varepsilon_C$. }
\end{figure}

Therefore, we consider the system sketched in Fig.~\ref{fig:3}(a), including two double-dot systems ($A$ and $B$), each internally coupled via a 
superconductor, and coupled to each other via another quantum dot ($C$).
By controlling the position of the energy level of dot $C$, $\varepsilon_C$, we can transfer an electron to or from dot $C$. If we do not measure which side 
($A$ or $B$) is involved in the charge transfer, we create entanglement between states where the charge transfer has flipped the parity of
system $A$ and states where it has flipped the parity of system $B$. The basic idea is related to Refs.~\onlinecite{Flensberg2011, Leijnse2011b, Leijnse12},
where, however, "standard" MBS were considered.

We illustrate the principle by showing how to both create maximally entangled parity states of $A$ and $B$, and  
measure the coherence time of such states. We consider the systems $A$ and $B$ to be tuned to the sweet spot,
where we can describe them in terms of the Majorana operators when investigating the low-energy physics. The total system is then described by 
the Hamiltonian
\begin{align}
\label{H_ABC}
  H_{ABC} &= \varepsilon_C n_C + t_{AC} \gamma_1^A \left( d_C - d_C^\dagger \right) +t_{BC} \gamma_1^B \left( d_C - d_C^\dagger \right),
\end{align}
where $n_C = d_C^\dagger d_C$ is the occupation operator for dot $C$. 
Without the coupling to dot $C$, the ground states of $A,B$ are $|e\rangle_{A,B}$ and $|o\rangle_{A,B}$,
where $|e,o\rangle_{A,B}$ stands for $|\alpha_{e,o}\rangle_{A,B}$ if $t^{A,B} < 0$ and for $|\beta_{e,o}\rangle_{A,B}$ if 
$t > 0$ [see Eq.~(\ref{sweetspotmany})].
We now consider the case $t_{AC} = t_{BC}$ (note that making the phases equal requires 
control of the phase difference, $\phi_A - \phi_B$, between the superconductors in $A$ and $B$). The total parity of the system 
($n_A + n_B + n_C$, with $n_{A,B} |e\rangle_{A,B} = 0$, $n_{A,B} |o \rangle_{A,B} = 1$) is conserved by the Hamiltonian~(\ref{H_ABC}) and 
we restrict our attention to the subspace of even total parity. Figure~\ref{fig:3}(b) shows the eigenenergies plotted as a function 
of $\varepsilon_C$. The special property of the limit $t_{AC} = t_{BC}$ is the crossing of two 
of the eigenstates (blue and magenta curves) at $\varepsilon_C = 0$. 
These states correspond to dot $C$ always being full (magenta sloped line) or empty (blue horisontal line), which we denote by 
$|F\rangle_{ABC}$ and $|E\rangle_{ABC}$, respectively. 
The other two states correspond to mixed occupation of dot $C$, denoted by $|M_1\rangle_{ABC}$ (brown lower line) and 
$|M_2\rangle_{ABC}$ (green upper line). 
When $\varepsilon_C \gg |t_{AC}|$, meaning far above the chemical potential of the superconductors, $|M_1\rangle_{ABC}$ ($|M_2\rangle_{ABC}$) 
corresponds to an empty (filled) dot, while the situation is reversed for $\varepsilon_C \ll -|t_{AC}|$.

We start with an empty dot $C$ ($n_C = 0$) and $\varepsilon_C \gg |t_{AC}|$, at the point marked $1$ in Fig.~\ref{fig:3}(b). 
We also initialize the systems $A$ and $B$ in the even parity states,
which can be done for example by moving the dot levels far above the chemical potential of the superconductors and waiting for the system 
to relax to the ground state, which is then even since the dots are empty and the superconductors have standard $s$-wave pairing. 
The initial state, $|i\rangle_{ABC}$, is then
\begin{align}
\label{initial}
  |i\rangle_{ABC} &= |e\rangle_A |e\rangle_B |0\rangle_C =  \frac{1}{\sqrt{2}} \left( |M_1\rangle_{ABC} + |E\rangle_{ABC} \right),
\end{align}
an equal superposition of the blue and brown states in Fig.~\ref{fig:3}(b), which are degenerate for $\varepsilon_C \gg |t_{AC}|$. 
We now adjust $V_g^C$ to bring down the level of dot $C$ to $\varepsilon_C \ll -|t_{AC}|$. If this is done adiabatically, the system will remain 
in an equal superposition of $|M_1\rangle_{ABC}$ and $|E\rangle_{ABC}$. However, this has now become a superposition of dot $C$ being 
empty [$|E\rangle_{ABC}$, corresponding to the point marked $2'$ in Fig.~\ref{fig:3}(b)] and full 
[$|M_1\rangle_{ABC}$, corresponding to the point marked $2$ in Fig.~\ref{fig:3}(b)], which will likely quickly decohere into a statistical mixture due to the long 
range of the Coulomb interaction. This is irrelevant for our purposes and if we desire, we can find out if the system is in $|E\rangle_{ABC}$
or $|M_1\rangle_{ABC}$ by measuring the charge on dot $C$. It is interesting to note that for $\varepsilon_C \ll -|t_{AC}|$ we have 
\begin{align}
\label{initial}
  |E\rangle_{ABC} &\rightarrow \frac{1}{\sqrt{2}} \left( |e\rangle_A |e\rangle_B  - |o\rangle_A |o\rangle_B \right), \\
  |M_1\rangle_{ABC} &\rightarrow \frac{1}{\sqrt{2}} \left( |e\rangle_A |o\rangle_B  + |o\rangle_A |e\rangle_B \right).
\end{align}
Thus, the adiabatic gate sweep may or may not result in dot $C$ being filled, but in any case prepares the systems $A$ and $B$ 
in a maximally entangled two-parity-qubit state.

We can now measure the coherence time, $T_2$, of the entangled states by waiting a time $\tau$ before making another 
adiabatic gate sweep back to $\varepsilon_C \gg |t_{AC}|$, after which the charge on dot $C$ is measured. If $\tau \ll T_2$, the system 
remains in either $|E\rangle_{ABC}$ or $|M_1\rangle_{ABC}$ and after the second gate sweep dot $C$ will always be empty 
[$n_C = 0$, point $3$ in Fig.~\ref{fig:3}(b)]. 
If, on the other hand, $\tau \gg T_2$, the system has time to decohere into a mixture of either $|E\rangle_{ABC}$ and $|M_2\rangle_{ABC}$,
or $|F\rangle_{ABC}$ and $|M_1\rangle_{ABC}$. In this case, after the second gate sweep dot $C$ will be empty or filled [point $3$ or $3'$ in Fig.~\ref{fig:3}(b)] 
with equal probabilities. The coherence time is found by many repeatitions of this measurement with different waiting times. 
This scheme can be described as parity to charge conversion, similar to spin to charge conversion used to measure coherence times 
in singlet-triplet qubits. \cite{Petta2005}

If $t_{AC} = t_{BC}$ is not perfectly fulfilled, a small avoided crossing appears between the states $|F\rangle_{ABC}$ and $|E\rangle_{ABC}$. 
The scheme described above works as long as we can make the gate sweep fast with respect to this avoided crossing, but slow with respect 
to the avoided crossing between $|M_1\rangle_{ABC}$ and $|M_2\rangle_{ABC}$.

\section{Conclusions}
In this paper we have introduced the concept of poor man's Majorana bound states, quasiparticle excitations which share all the characteristics 
of "standard" Majorana bound states, but lack topological protection. The poor man's Majoranas form in a rather simple setup consisting of two 
quantum dots coupled via a standard $s$-wave superconductor and placed in an inhomogeneous magnetic field. 
Under the appropriate conditions, two spatially separated MBS appear, one on each dot, as can be verified by tunnel spectroscopy. 
The fermionic parity qubit formed by the two Majoranas is non-local and cannot be measured by probing one dot only. 
We believe that the suggested system is a very experimentally attractive platform in which to test some of the exotic Majorana 
physics which has been suggested theoretically.

We have also discussed coupling of two parity qubits via an additional quantum dot. This setup allows entanglement of parity qubits 
through gate-controlled charge transfer, as well as measurements of the associated coherence times. 

While finalizing the manuscript we became aware of the somewhat related work \cite{otherdots}.

\section{Acknowledgements}
We thank C. M. Marcus for valuable discussions and feedback.

\bibliographystyle{/usr/share/texmf-texlive/bibtex/bst/revtex/apsrev4-1.bst}
\end{document}